\documentclass{elsart}
\usepackage{graphicx}

\begin{document}
\begin{frontmatter}

  \title{Molecular dynamics for full QCD
    simulations with an improved action}

  \author[ZSU,HLRZ,DESY]{Xiang-Qian Luo}

\address[ZSU]{Department of Physics, Zhongshan University, Guangzhou 510275, China\thanksref{Mail}}

\address[HLRZ]{HLRZ (Supercomputing Center), Forschungszentrum, D-52425 J\"ulich, Germany}

\address[DESY]{Deutsches Elektronen-Synchrotron DESY, D-22603 Hamburg, Germany}

\thanks[Mail]{Present Address. E-Mail:  stslxq@zsu.edu.cn}

\begin{abstract}
  I derive the equation of motion in molecular dynamics
  for doing full lattice QCD  simulations 
  with clover quarks. The even-odd preconditioning technique,
   expected to significantly reduce the computational effort,
  is further developed 
for the simulations. 
\end{abstract}


\end{frontmatter}

Published in {\it Computer Physics Communications} {\bf 94} (1996) 119-127.

\section{Introduction}
The effects of dynamical quarks are important in QCD at finite
temperature as
well as in some phenomenological aspects at zero temperature.
Unfortunately, the inclusion of dynamical quarks is the most demanding task
in computer simulations of lattice QCD. 
The hybrid molecular dynamics or Hybrid Monte Carlo (HMC)
methods have been developed into very efficient algorithms
(maybe the most popular) for dynamical
quarks. In these algorithms,
the equation of motion is the essential ingredient.
One has to derive the relevant equations before writing the
programs for molecular dynamical simulations.
For lattice QCD with staggered or Wilson fermions, these equations have
already been available in the literature \cite{Gottlieb,Gupta,Sarno}.

Lattice \rm{QCD} has discretization
errors due to the lattice spacing $a$. At intermediate bare coupling,
corresponding to relatively large $a$,
these systematic errors might sometimes be
very severe for Wilson fermions due to the chiral symmetry breaking term.
The current computers do not allow the calculations done for very small
$a$, because to reduce $a$ implies to use a much larger
lattice. Another way out is to use the improved fermionic actions.
Recently, it has been shown that the use of the clover action
\cite{SW,Sach}
can significantly reduce these finite cut-off errors.
However, the calculations of the clover action are much more complicated
than
the standard Wilson action.
To my knowledge, there has not been a simulation of lattice QCD with the
clover action in the literature.

The purpose of this paper is to derive the equation of motion
for full \rm{QCD} simulations with the clover action. 
Because the fermionic matrix has to be inverted in each step of
the molecular
dynamics step,
it is also challenging to devise efficient algorithms for
preconditioning
\cite{Gupta,De} the
fermionic matrix so that the inverse is easier to compute.
For this reason, I also
extend the even-odd preconditioning technique, previously used
for quark
propagator measurements, to the case of dynamical clover fermions.

\section{Preconditioning}

\subsection{The action}
The action of the theory is $S=S_G +S_F$,
where 
\begin{eqnarray*}
S_G=-
{\beta \over N_c} \sum_{p} Re ~tr (U_p)
\end{eqnarray*}
\begin{eqnarray}
=
- {\beta \over N_c} \sum_{x,\mu >\nu}
Re ~tr \lbrack U_{\mu}(x)U_{\nu}(x+\mu)
U_{\mu}(x+\nu)^{\dagger}U_{\nu}(x)^{\dagger} \rbrack
\end{eqnarray}
is the gauge action.
The clover action for the quarks \cite{SW,Sach}  is
\begin{eqnarray*} 
  S_F= \sum_{x,y} {\bar \psi}(x) M_{xy} \psi (y)
  =  \sum_{x,y} {\bar \psi}(x)(A_{x,y} -\kappa B_{x,y})\psi (y),
\end{eqnarray*}
\begin{eqnarray*} 
  A_{x,y} =\delta_{x,y}\lbrack1
  - {\kappa C \over 2} \sum_{\mu, \nu} \sigma_{\mu \nu}
{\mathcal   F_{\mu \nu}}(x)\rbrack,
\end{eqnarray*}
\begin{eqnarray} 
  B_{x,y} = \sum_{\mu=1}^4 (1-\gamma_{\mu}) U_{\mu}(x) \delta_{x,y-\mu}+
  (1+\gamma_{\mu}) U^{\dagger}_{\mu}(x-\mu)  \delta_{x,y+\mu}.
  \label{sf}
\end{eqnarray}
$A$ is local and hermitian, and $B$ connects only the nearest
neighbor sites.
The field strength tensor on the lattice is defined by
$ {\mathcal F_{\mu \nu}}(x)=
\lbrack Q_{\mu \nu}(x) - Q_{\mu \nu}^{\dagger}(x)\rbrack/ 2i$,
where $Q_{\mu \nu}$ is the averaged
sum of four plaquettes on the $\mu \nu$ plane
with the lattice site $x$ as one corner. Each plaquette is
the product of four link variables in the counterclockwise sense
and begins with the link variable directed away from the site $x$ and
ends with the link variable directed towards site $x$, i.e.,
\begin{eqnarray*} 
  Q_{\mu \nu}(x)
  ={1 \over 4} \lbrack U_{\mu}(x)U_{\nu}(x+\mu)
  U_{\mu}(x+\nu)^{\dagger}U_{\nu}(x)^{\dagger}
\end{eqnarray*}
\begin{eqnarray*} 
  +U_{\nu}(x)U_{\mu}(x-\mu+\nu)^{\dagger}
  U_{\nu}(x-\mu)^{\dagger}U_{\mu}(x-\mu)
\end{eqnarray*}
\begin{eqnarray*} 
+U_{\mu}(x-\mu)^{\dagger} U_{\nu}(x-\mu-\nu)^{\dagger}
U_{\mu}(x-\mu-\nu)U_{\nu}(x-\nu)
\end{eqnarray*}
\begin{eqnarray} 
  +U_{\nu}(x-\nu)^{\dagger} U_{\mu}(x-\nu)
  U_{\nu}(x+\mu-\nu)U_{\mu}(x)^{\dagger}\rbrack
\end{eqnarray}
as shown in Fig. 1.
This operator is so chosen as for the maximal symmetry on the lattice. 
Most symbols in above equations are conventional,
while the coefficient $C$ in
(\ref{sf}) depends on the choice of improvement strategy:
$C=1$ for tree level
improvement, and $C=\lbrack Re ~tr (U_p)/N_c\rbrack ^{-3/4}$
for the tadpole improvement \cite{LM}.

\begin{figure}
\begin{center}
\rotatebox{0}{\includegraphics[width=10cm]{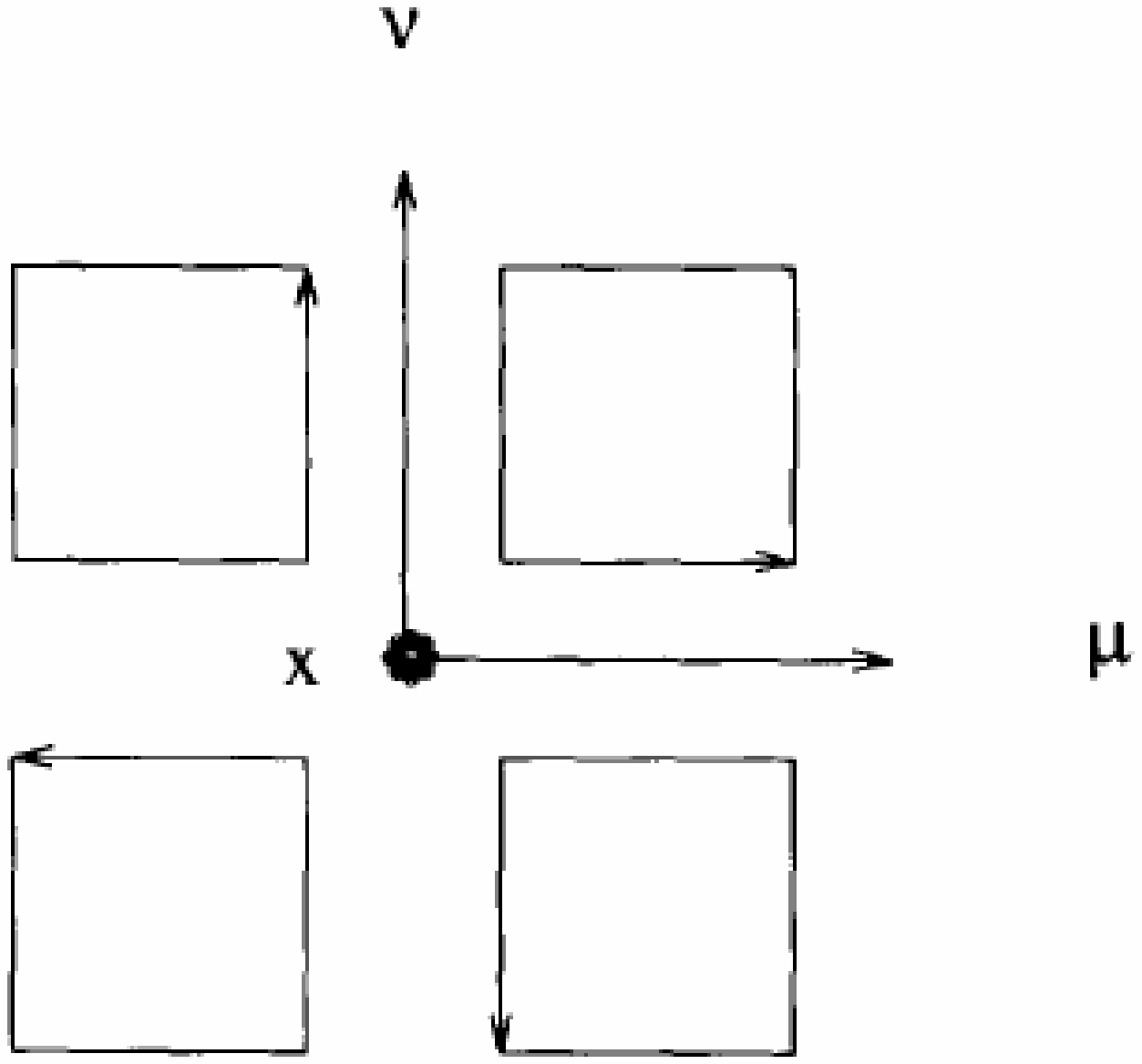}}
 \end{center}
\vspace{-7cm}
\caption{ 
Plot of the clover operator $Q_{\mu \nu}$. The product in the
plaquette is in counterclockwise sense
and begins with the directed link.}
\vspace{1cm}
\end{figure}

\subsection{Even-odd splitting}
The lattice sites can be organized in an
even-odd checkerboard and the even
sites are numbered before the odd sites
such that the fermionic matrix
can be written as
\cite{UKQCD,Stanford}
\begin{eqnarray}
  M =\left( 
    \begin{array}{cc}
     A_{ee} & - \kappa B_{eo} \\
     - \kappa B_{oe}              & A_{oo} 
     \end{array}
   \right),
\label{decomp}
\end{eqnarray}
where $e$ or $o$ denotes even or odd site on the lattice.

Using such an arrangement, we obtain
\begin{eqnarray}
  \det(M) =\det (A_{oo})~ \det(A_{ee} -\kappa^2 B_{eo} A^{-1}_{oo} B_{oe})
  = \det A_{oo} ~\det M_{ee}
\end{eqnarray}
where
\begin{eqnarray}
  M_{ee}=A_{ee} -\kappa^2 B_{eo} A^{-1}_{oo} B_{oe},
  \label{MEE}
\end{eqnarray}
which couples only to even sites of the lattice. Now  $\det(M)$
on the whole lattice has been factorized as a product of
the determinant of the local
matrix $A$ on the odd lattice and that of $M_{ee}$ on the even lattice.

To calculate the fermionic determinant, it is
useful to introduce the pseudo-scalar variables $\eta_o$ and $\phi_e$,
so that
\begin{eqnarray*}
  \det (M^{\dagger} M) = \det (A^{\dagger}_{oo} A_{oo})
  ~\det (M^{\dagger}_{ee} M_{ee})
  =\int d \eta_o^{\dagger}~ d \eta_o \int d \phi_e^{\dagger}~
  d \phi_e exp(-S_{pf})
\end{eqnarray*}
\begin{eqnarray}
  =  \int d \eta_o^{\dagger}~ d \eta_o  ~
  exp\lbrack-\eta_o^{\dagger}(A^{\dagger}_{oo}
  A_{oo})^{-1} \eta_o\rbrack
  \int d \phi_e^{\dagger}~ d \phi_e ~
  exp\lbrack-\phi_e^{\dagger}(M^{\dagger}_{ee}
  M_{ee})^{-1} \phi_e\rbrack,
\end{eqnarray}
where $S_{pf}$ is the pseudo-fermionic action
\begin{eqnarray}
S_{pf}=\ \eta_o^{\dagger}(A^{\dagger}_{oo}
A_{oo})^{-1} \eta_o
+  \phi_e^{\dagger}(M^{\dagger}_{ee}
M_{ee})^{-1} \phi_e,
\end{eqnarray}
describing two flavor quarks with the same bare mass. Notice that
$\eta_o=A^{\dagger}_{oo} \xi_o$ and $\phi_e=M^{\dagger}_{ee} \theta_e$
have no direct coupling, where $\xi_o$ and $\theta_e$
are Gaussian noises injected at beginning of
each molecular dynamics trajectory and held fixed during
each trajectory.

In the remaining text, I will use the above even-odd preconditioning to
discuss
the molecular dynamics.

\subsection{Fermionic inversion}
For quark propagator measurements and also
in each molecular dynamics step,
one has to calculate $M_{ee}^{-1}$ or
$(M^{\dagger}_{ee} M_{ee})^{-1}$, which can be implemented using the
standard techniques like minimum residue,
conjugate gradient or stabilized
biconjugate gradient algorithms. The advantage of the even-odd splitting,
as can also be seen later, is that such inversion
is implemented only on the
even lattice. Furthermore, due to the factor $\kappa^2$ in (\ref{MEE}),
$M_{ee}$ is better conditioned than $M$.

For the inversion $A_{oo}^{-1}$ on each odd site,
because it is completely local,
we can use the $LDL^{\dagger}$ decomposition
\cite{Stanford,Linear} to solve it.
Since it is a hermitian
matrix,
there exists a diagonal matrix $D$ and lower-triangular matrix $L$ such that
$A=LDL^{\dagger}$.
Denoting $i$ and $j$
as the color-spin indexes of $A$, then
\begin{eqnarray*}
  D_{i}=A_{ii}-\sum_{k=1}^{i-1}L_{ik}D_{k}L^{\ast}_{ik}, ~ ~
\end{eqnarray*}
\begin{eqnarray}
  L_{ij}D_j=A_{ij}-\sum_{k=1}^{j-1}L_{ik}D_{k}L^{\ast}_{jk},
  ~ ~ (j=1,...,i-1).
\end{eqnarray}
We can also compute the solution of $AX=b$ by
$y=L^{-1}b$ and $X=(L^{\dagger})^{-1}D^{-1}Y$, i.e.,
\begin{eqnarray*}
  Y_{i}=b_{i}-\sum_{k=1}^{i-1}L_{ik}Y_{k}, ~ ~    (i=1,...,n), ~ ~
\end{eqnarray*}
\begin{eqnarray}
  X_{i}=Y_{i}/D_i-\sum_{k=i+1}^{n}L_{ki}^{\ast}X_{k}, ~ ~ (i=n,...,1),
\end{eqnarray}
with $n=12$, which is the number of colors times the number of spins.
The calculation is quite easy because there are $n^2$ multiplications
and only $n$ divisions.

\section{Molecular dynamics}
\subsection{Equation of motion}
To develop the equation of motion for the gauge field 
${\mathcal A_{\mu}}(x)$, one has 
to introduce a Hamiltonian 
$H=
\sum_{x,\mu} P^2_{{\mathcal A_{\mu}}(x)}/2 +S_g+S_{pf}$,
with $P_{{\mathcal A_{\mu}}(x)}$ 
the canonical conjugate momentum defined by
$P_{{\mathcal A_{\mu}}(x)}= \partial H
/ \partial (d {\mathcal A_{\mu}}(x)/d\tau)$,
and $\tau$ being the fictious molecular dynamics time.

The gauge configurations are generated
by solving the Hamiltonian equation of motion:
\begin{eqnarray}
  {d {\mathcal A_{\mu}}(x)\over d\tau}=P_{{\mathcal  A_{\mu}}(x)},
  ~ ~
{d P_{{\mathcal A_{\mu}}(x)} \over d\tau}=
{\partial H \over \partial {\mathcal A_{\mu}}(x)}=
-{\partial S_G
  \over \partial {\mathcal A_{\mu}}(x)}
-{\partial S_{pf} \over \partial {\mathcal A_{\mu}}(x)}. 
\label{mol}
\end{eqnarray} 
For the gauge action, it is quite easy to show that  
\begin{eqnarray*}
-{\partial S_G
  \over \partial {\mathcal A_{\mu}}(x)}
={\beta \over 2 N_c}
\lbrack {\partial U_{\mu}(x)
  \over \partial {\mathcal A_{\mu}}(x)} STAPLE_{\mu}(x)
+ STAPLE_{\mu}(x)^{\dagger} {\partial U_{\mu}(x)^{\dagger}
  \over \partial {\mathcal A_{\mu}}(x)}\rbrack
\end{eqnarray*}
\begin{eqnarray}
  ={i \beta \over 2 N_c}
  \lbrack U_{\mu}(x) STAPLE_{\mu}(x) -h.c.\rbrack,
  \label{dsg}
\end{eqnarray}
where $STAPLE_{\mu}(x)$ is the sum over six staples
surrounding the link $U_{\mu}(x)$.
For the fermionic part, 
\begin{eqnarray}
 - {\partial S_{pf} \over \partial {\mathcal A_{\mu}}(x) }=
  X_o^{A \dagger} {\partial (A^{\dagger}_{oo}
     A_{oo}) \over  \partial {\mathcal A_{\mu}}(x)} X_o^{A}
 + X_e^{\dagger} {\partial (M^{\dagger}_{ee}
     M_{ee}) \over  \partial {\mathcal A_{\mu}}(x)} X_e,
\label{dsf1}
 \end{eqnarray}
 where 
\begin{eqnarray}
X_o^A=(A^{\dagger}_{oo}
A_{oo})^{-1} \eta_o, ~ ~
X_e= (M^{\dagger}_{ee}
M_{ee})^{-1} \phi_e.
\label{XY1}
 \end{eqnarray}
If we define two more variables
 \begin{eqnarray}
  Y_o^{A}=A_{oo}X_o^A, ~ ~
  Y_e=M_{ee} X_e,
\label{XY2}
\end{eqnarray}
(\ref{dsf1}) can be simply rewritten as
 \begin{eqnarray*}
   - {\partial S_{pf} \over \partial {\mathcal A_{\mu}}(x) }
   =X_o^{A \dagger} {\partial A^{\dagger}_{oo} \over  \partial {\mathcal
    A_{\mu}}(x)} Y_o^{A} +
Y_o^{A \dagger} {\partial A_{oo} \over  \partial {\mathcal
    A_{\mu}}(x)} X_o^A
\end{eqnarray*}
\begin{eqnarray}
+ X_e^{\dagger} {\partial M^{\dagger}_{ee} \over  \partial {\mathcal
    A_{\mu}}(x)} Y_e +
Y_e^{\dagger} {\partial M_{ee} \over  \partial
  {\mathcal  A_{\mu}}(x)} X_e.
\label{dsf}
\end{eqnarray}

A straightforward computation leads to
  \begin{eqnarray*}
    {\partial M_{ee} \over \partial {\mathcal A_{\mu}}(x)}=
    {\partial A_{ee} \over \partial {\mathcal A_{\mu}}(x)}
    -\kappa^2 {\partial B_{eo} \over \partial {\mathcal A_{\mu}}(x)}A_{oo}^{-1}
    B_{oe}
\end{eqnarray*}
\begin{eqnarray*} 
  +\kappa^2 B_{eo} A_{oo}^{-1} {\partial A_{oo}
    \over \partial {\mathcal A_{\mu}}(x)}
    A_{oo}^{-1} B_{oe}
    - \kappa^2  B_{oe} A_{oo}^{-1} {\partial B_{oe}
      \over \partial {\mathcal A_{\mu}}(x)},
  \end{eqnarray*}
    \begin{eqnarray*}
    {\partial M^{\dagger}_{ee} \over \partial {\mathcal A_{\mu}}(x)}=
    {\partial A_{ee} \over \partial {\mathcal A_{\mu}}(x)}
    -\kappa^2 {\partial B^{\dagger}_{eo} \over \partial
      {\mathcal A_{\mu}}(x)}A_{oo}^{-1}
    B^{\dagger}_{oe}
\end{eqnarray*}
\begin{eqnarray} 
  +\kappa^2 B^{\dagger}_{eo} A_{oo}^{-1} {\partial A_{oo}
    \over \partial {\mathcal A_{\mu}}(x)}
    A_{oo}^{-1} B^{\dagger}_{oe}
    - \kappa^2  B^{\dagger}_{oe} A_{oo}^{-1}
    {\partial B^{\dagger}_{oe} \over \partial {\mathcal A_{\mu}}(x)}.
  \end{eqnarray}
By defining the following variables on the odd sites
\begin{eqnarray}
X_o = \kappa A_{oo}^{-1} B_{oe} X_e, ~ ~
Y_o = \kappa A_{oo}^{-1} B_{oe}^{\dagger}Y_e,
\label{XY3}
  \end{eqnarray}
we have
\begin{eqnarray*}
 Y_e^{\dagger} {\partial 
   M_{ee} \over  \partial {\mathcal A_{\mu}}(x)} X_e=
 Y_e^{\dagger}{\partial A_{ee} \over \partial {\mathcal A_{\mu}}(x)}X_e
 +Y_o^{\dagger} {\partial A_{oo} \over \partial {\mathcal A_{\mu}}(x)}X_o
\end{eqnarray*}
\begin{eqnarray*} 
 -\kappa Y_e^{\dagger}{\partial B_{eo} \over \partial
   {\mathcal A_{\mu}}(x)} X_o
 -\kappa Y_o^{\dagger}{\partial B_{oe} \over \partial
   {\mathcal A_{\mu}}(x)} X_e,
 \end{eqnarray*}
\begin{eqnarray*}
 X_e^{\dagger} {\partial 
   M_{ee}^{\dagger} \over  \partial {\mathcal A_{\mu}}(x)} Y_e=
 X_e^{\dagger}{\partial A_{ee} \over \partial {\mathcal A_{\mu}}(x)}Y_e
 +X_o^{\dagger} {\partial A_{oo} \over \partial {\mathcal A_{\mu}}(x)}Y_o
\end{eqnarray*}
\begin{eqnarray} 
 -\kappa X_e^{\dagger}{\partial B^{\dagger}_{eo} \over \partial
   {\mathcal A_{\mu}}(x)} Y_o
 -\kappa X_o^{\dagger}{\partial B^{\dagger}_{oe} \over \partial
   {\mathcal A_{\mu}}(x)} Y_e.
 \label{DM}
\end{eqnarray}
We can further demonstrate that the last two terms of these two
equations in (\ref{DM}) are summarized as
\begin{eqnarray*}
 -\kappa Y_e^{\dagger}{\partial B_{eo} \over \partial
   {\mathcal A_{\mu}}(x)} X_o
 -\kappa Y_o^{\dagger}{\partial B_{oe} \over \partial
   {\mathcal A_{\mu}}(x)} X_e
 -\kappa X_e^{\dagger}{\partial B^{\dagger}_{eo} \over \partial
   {\mathcal A_{\mu}}(x)} Y_o
 -\kappa X_o^{\dagger}{\partial B^{\dagger}_{oe} \over \partial
   {\mathcal A_{\mu}}(x)} Y_e
 \end{eqnarray*}
\begin{eqnarray}
  =-\kappa \sum_{x', y} \lbrack Y_{x'}^{\dagger}{\partial B_{x'y}
    \over \partial  {\mathcal A_{\mu}}(x)} X_y +
 X_{x'}^{\dagger}{\partial B^{\dagger}_{x'y} \over \partial
   {\mathcal A_{\mu}}(x)} Y_y\rbrack
 =- i \kappa \lbrack U_{\mu}(x) F_{\mu}^W -h.c.\rbrack,
\label{DB}
\end{eqnarray}
where
\begin{eqnarray}
  F_{\mu}^W=tr_{dirac} \lbrack(1+\gamma_{\mu}) Y_{x+\mu} X_x^{\dagger} 
  + (1-\gamma_{\mu})X_{x+\mu} Y_x^{\dagger}\rbrack,
\end{eqnarray}
the same form as the fermionic force in the Wilson fermion case.  
As can also be seen later, the last two terms
in (\ref{dsf}) have exactly the
same form for even and odd sites.
Therefore, the introduction of the variables (\ref{XY1}),
(\ref{XY2}) and
(\ref{XY3}) has a great advantage. 

A remark has to be made: to keep the conjugate momentum traceless,
the right hand side (r.h.s.) of
 (\ref{mol})
 should finally be subtracted by a term being the trace of the r.h.s.
 divided by $N_c$.

 \subsection{Practical implementation}

 The simulations should be carried out in the first two steps as follows.

 \noindent
 1) Generating the full configurations.
 In numerical integration of the equation
 of motion, one has to Taylor expand
 $U(\tau+d\tau)=exp\lbrack i d \tau P(\tau)\rbrack U(\tau)
 =U(\tau)+i d \tau P(\tau)U(\tau) +...$,
 $P(\tau+d \tau)= P(\tau)+ d\tau {dP \over d \tau} (\tau)+ ...$,
 with finite order truncation.
 The leapfrog scheme can reduce the truncation errors to
 $N_{md}O(d \tau^3)=O(d \tau^2)$ at $N_{md}$ molecular dynamical steps.
 These errors can be canceled
 by a Metropolis test at the end of the trajectory.
 Of course, this $d \tau$
 has to be fine tuned to maintain high acceptance
 rate and small auto-correlation time.

 \noindent
2) About the clover coefficient.  
 For the tree level improvement, $C=1$.
 For the tadpole improvement scheme,
 the value of $C$ depends dynamically on the
 configurations.  One has to
determine $C$ self-consistently from the simulation. 
For example, one may first have an initial guess for it,
then generate a gauge configuration.
From the plaquette, we can get a new $C$ value.
After some iteration, $C$
might
converge to some stable value for some given $\beta$ and $\kappa$.
(This could  be done
very quickly since the plaquette can be accurately measured with a
small number of configurations and on small lattices,
provided the system is far
away from a phase transition). 

 \noindent
 3) Measuring the physical quantities.
To obtain the improved  hadronic matrix
 elements, 
 rotation of quark fields \cite{Sach,UKQCD} is necessary.

 \section{More details about the fermionic force}

\begin{figure}
\begin{center}
\rotatebox{0}{\includegraphics[width=10cm]{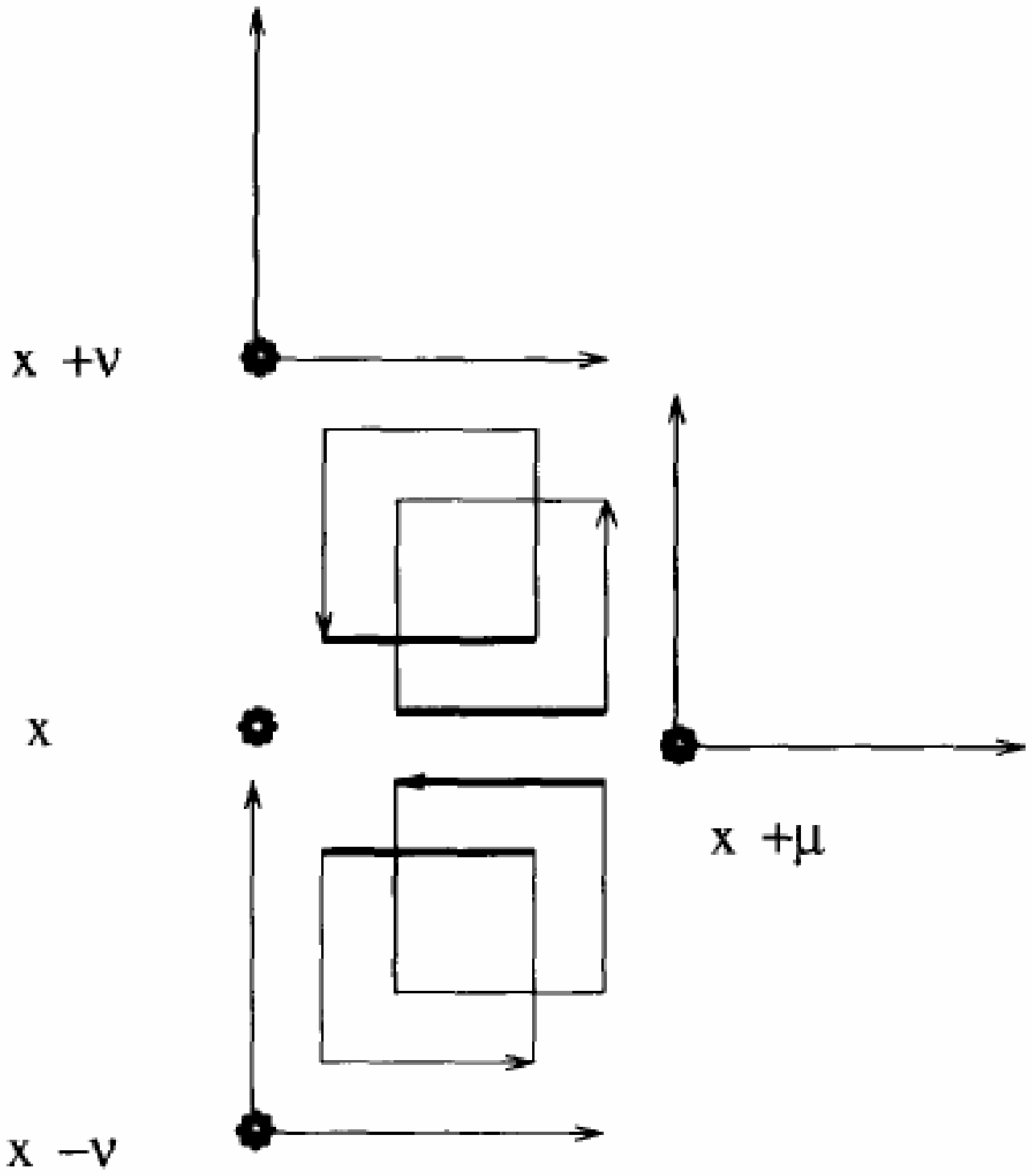}}
 \end{center}
\vspace{-1cm}
\caption{ 
Relevant plaquettes for
$\sum_y \lbrack Y_y^{A \dagger}
(\partial A_{yy} /\partial {\mathcal  A_{\mu}}(x))
X_y^{A}+
X_y^{A \dagger} (\partial A^{\dagger}_{yy} /\partial {\mathcal  A_{\mu}}(x))
Y_y^{A}\rbrack$
when $x \in y$, where the thick lines indicate the links relevant for
$\partial U_{\mu}(x)/ {\partial \mathcal A_{\mu}}(x)$ or 
$\partial U_{\mu}(x)^{\dagger}/ {\partial \mathcal A_{\mu}}(x)$.}
\vspace{1cm}
\end{figure}

I have described that how the introduction of  the variables (\ref{XY1}),
(\ref{XY2}) and
(\ref{XY3}) leads to the equation of motion similar
to that of the Wilson case.
What is different is the terms with matrix $A$,
which makes the equation of
motion much more complicate. Therefore, it deserves further discussions.
Note the term in the pseudo-fermion action
$Y_o^{A \dagger} A_{oo} X_o^A$ (also the second term)
is placed only on odd sites of the lattice,
then for $x$ being odd sites,
there are terms only on $x$, $x+\mu+\nu$ and $x+\mu-\nu$
relevant for
$Y_o^{A \dagger} (\partial A_{oo} /\partial {\mathcal  A_{\mu}}(x)) X_o^A$
as shown in Fig. 2, i.e.,

\begin{eqnarray*}
 {1 \over 2}  Y_o^{A \dagger} \sum_{\mu' \not= \nu}
    \sigma_{\mu' \nu} {\partial Q_{\mu' \nu}(o) \over \partial
    {\mathcal  A_{\mu}}(x)} X_o^A=
\end{eqnarray*}
\begin{eqnarray*}
  {1 \over 4} \sum_{\nu} ~'~ Y_{x}^{A \dagger} \sigma_{\mu \nu}
  \lbrack {\partial U_{\mu}(x) \over {\partial \mathcal
      A_{\mu}}(x)}U_{\nu}(x+\mu)U_{\mu}(x+\nu)^{\dagger}
  U_{\nu}(x)^{\dagger}
  \end{eqnarray*}
\begin{eqnarray*}
+U_{\nu}(x-\nu)^{\dagger} U_{\mu}(x-\nu) U_{\nu}(x+\mu-\nu){\partial
  U_{\mu}(x)^{\dagger}\over {\partial \mathcal A_{\mu}}(x)}\rbrack X_x^{A}
\end{eqnarray*}
\begin{eqnarray*} 
  + {1 \over 4} \sum_{\nu} ~'~
  Y_{x+\mu+\nu}^{A \dagger} \sigma_{\mu \nu}
  U_{\mu}(x+\nu)^{\dagger}U_{\nu}(x)^{\dagger}{\partial U_{\mu}(x)
    \over {\partial \mathcal  A_{\mu}}(x)}U_{\nu}(x+\mu) X_{x+\mu+\nu}^{A}
\end{eqnarray*}
\begin{eqnarray*} 
+ {1 \over 4} \sum_{\nu} ~'~ Y_{x+\mu-\nu}^{A \dagger} \sigma_{\mu \nu}
U_{\nu}(x+\mu-\nu){\partial U_{\mu}(x)^{\dagger}
  \over {\partial \mathcal A_{\mu}}(x)}
U_{\nu}(x-\nu)^{\dagger} U_{\mu}(x-\nu) X_{x+\mu-\nu}^{A}.
\end{eqnarray*}
\begin{eqnarray}
  ~
\end{eqnarray}

Here $\sum_{\nu}'$ means the sum over $\nu \ne \mu$.
For $x$ being even sites, there are terms only on $x+\mu$,
$x+\nu$ and $x-\nu$
as shown in Fig. 3, i.e.,

\begin{eqnarray*}
  {1 \over 2} Y_o^{A \dagger} \sum_{\mu' \not= \nu}
  \sigma_{\mu' \nu} {\partial Q_{\mu' \nu}(o) \over {\partial
      \mathcal  A_{\mu}}(x)} X_o^A=
\end{eqnarray*}
\begin{eqnarray*}
  {1 \over 4} \sum_{\nu} ~'~ Y_{x+\mu}^{A \dagger} \sigma_{\mu \nu}
  \lbrack U_{\nu}(x+\mu)U_{\mu}(x+\nu)^{\dagger}U_{\nu}(x)^{\dagger}
  {\partial U_{\mu}(x) \over {\partial \mathcal  A_{\mu}}(x)}
  \end{eqnarray*}
\begin{eqnarray*}
+ {\partial
   U_{\mu}(x)^{\dagger}\over {\partial \mathcal A_{\mu}}(x)}
 U_{\nu}(x-\nu)^{\dagger} U_{\mu}(x-\nu)
 U_{\nu}(x+\mu-\nu)\rbrack X_{x+\mu}^{A}
  \end{eqnarray*}
\begin{eqnarray*}
+{1 \over 4} \sum_{\nu} ~'~ Y_{x+\nu}^{A \dagger} \sigma_{\mu \nu}
U_{\nu}(x)^{\dagger}
{\partial U_{\mu}(x) \over {\partial \mathcal  A_{\mu}}(x)}
U_{\nu}(x+\mu)U_{\mu}(x+\nu)^{\dagger}X_{x+\nu}^{A}
  \end{eqnarray*}
\begin{eqnarray*}
+{1 \over 4} \sum_{\nu} ~'~ Y_{x-\nu}^{A \dagger} \sigma_{\mu \nu}
U_{\mu}(x-\nu) U_{\nu}(x+\mu-\nu){\partial U_{\mu}(x)^{\dagger}
  \over {\partial \mathcal A_{\mu}}(x)}
U_{\nu}(x-\nu)^{\dagger}X_{x-\nu}^{A}.
  \end{eqnarray*}
\begin{eqnarray}
~
\end{eqnarray}

\begin{figure}
\begin{center}
\rotatebox{0}{\includegraphics[width=10cm]{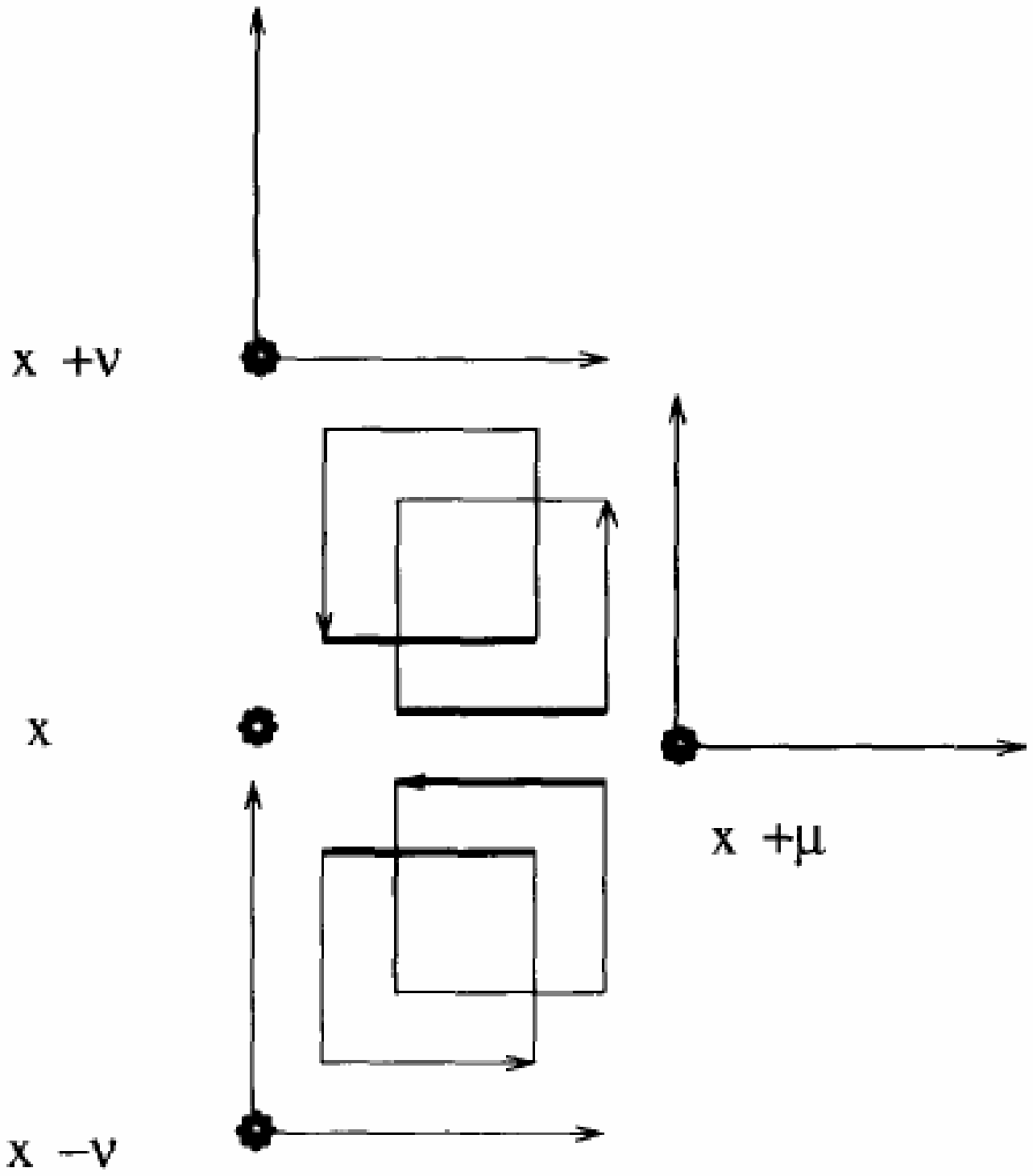}}
 \end{center}
\vspace{-5cm}
\caption{The same as Fig. 2 but for $x$ doesn't belong to $y$.}
\vspace{1cm}
\end{figure}

Therefore for odd-site $x$, the first two terms in (\ref{dsf}) read
\begin{eqnarray*}
  Y_o^{A \dagger} {\partial A_{oo} \over  {\partial
      \mathcal  A_{\mu}}(x)} X_o^A
  + X_{o}^{A \dagger} {\partial A_{oo}^{\dagger} \over  {\partial
      \mathcal  A_{\mu}}(x)} Y_o^{A}=
  \end{eqnarray*}
\begin{eqnarray*}
  - {\kappa C \over 8}\lbrack  \sum_{\nu} ~'~
  Y_{x}^{A \dagger} \sigma_{\mu \nu}
 {\bf U_{\mu}(x)} U_{\nu}(x+\mu)
 U_{\mu}(x+\nu)^{\dagger}U_{\nu}(x)^{\dagger}X_x^{A}
  \end{eqnarray*}
\begin{eqnarray*}
  - \sum_{\nu} ~'~ Y_{x}^{A \dagger} \sigma_{\mu \nu}
  U_{\nu}(x-\nu)^{\dagger} U_{\mu}(x-\nu) U_{\nu}(x+\mu-\nu)
  {\bf U_{\mu}(x)}^{\dagger}  X_x^{A}
\end{eqnarray*}
\begin{eqnarray*} 
  + \sum_{\nu} ~'~ Y_{x+\mu+\nu}^{A \dagger} \sigma_{\mu \nu}
  U_{\mu}(x+\nu)^{\dagger}U_{\nu}(x)^{\dagger}
  {\bf U_{\mu}(x)} U_{\nu}(x+\mu)
  X_{x+\mu+\nu}^{A}
\end{eqnarray*}
\begin{eqnarray*} 
  -\sum_{\nu} ~'~ Y_{x+\mu-\nu}^{A \dagger} \sigma_{\mu \nu}
  U_{\nu}(x+\mu-\nu)
  {\bf U_{\mu}(x)}^{\dagger}
U_{\nu}(x-\nu)^{\dagger} U_{\mu}(x-\nu)
X_{x+\mu-\nu}^{A}
\end{eqnarray*}
\begin{eqnarray} 
  +(Y \leftrightarrow X) +h.c.\rbrack.
  \label{DAO}
\end{eqnarray}

 Similarly, for even $x$, the first two terms in (\ref{dsf}) are
\begin{eqnarray*}
  Y_o^{A \dagger} {\partial A_{oo} \over  {\partial
      \mathcal  A_{\mu}}(x)} X_o^A
  + X_{o}^{A \dagger} {\partial A_{oo}^{\dagger} \over  {\partial
      \mathcal  A_{\mu}}(x)} Y_o^{A}=
  \end{eqnarray*}
\begin{eqnarray*}
  - {\kappa C \over 8}\lbrack\sum_{\nu} ~'~
  Y_{x+\mu}^{A \dagger} \sigma_{\mu \nu}
  U_{\nu}(x+\mu)U_{\mu}(x+\nu)^{\dagger}U_{\nu}(x)^{\dagger}
  {\bf U_{\mu}(x)} X_{x+\mu}^{A}
  \end{eqnarray*}
\begin{eqnarray*}
  -  \sum_{\nu} ~'~ Y_{x+\mu}^{A \dagger} \sigma_{\mu \nu}
  {\bf U_{\mu}(x)}^{\dagger}
 U_{\nu}(x-\nu)^{\dagger} U_{\mu}(x-\nu)
 U_{\nu}(x+\mu-\nu)X_{x+\mu}^{A}
 \end{eqnarray*}
\begin{eqnarray*}
+ \sum_{\nu} ~'~ Y_{x+\nu}^{A \dagger} \sigma_{\mu \nu}
U_{\nu}(x)^{\dagger}
{\bf U_{\mu}(x)}
U_{\nu}(x+\mu)U_{\mu}(x+\nu)^{\dagger}X_{x+\nu}^{A}
  \end{eqnarray*}
\begin{eqnarray*}
- \sum_{\nu} ~'~ Y_{x-\nu}^{A \dagger} \sigma_{\mu \nu}
U_{\mu}(x-\nu) U_{\nu}(x+\mu-\nu){\bf U_{\mu}(x)}^{\dagger}
U_{\nu}(x-\nu)^{\dagger}
X_{x-\nu}^{A}
\end{eqnarray*}
\begin{eqnarray} 
  +(Y \leftrightarrow X) +h.c.\rbrack.
  \label{DAE}
\end{eqnarray}

Equations (\ref{DAO}) and (\ref{DAE})
can be generalized to terms with  $y$ being odd or even:

\begin{eqnarray*} 
\sum_y  \lbrack Y_y^{\dagger} {\partial A_{yy} \over  {\partial
      \mathcal  A_{\mu}}(x)} X_y
  + X_y^{\dagger} {\partial A_{yy}^{\dagger} \over  {\partial
      \mathcal  A_{\mu}}(x)} Y_y\rbrack
  \end{eqnarray*}
\begin{eqnarray*}
  =(\ref{DAO}), ~ ~ if ~  x \in y,
 \end{eqnarray*}
\begin{eqnarray} 
  =(\ref{DAE}), ~ ~ if ~  x \notin y.
  \label{general}
\end{eqnarray}
These relations are also quite useful when deriving the
first two terms in (\ref{DM}). 

Summarizing (\ref{DB}) and (\ref{general}), (\ref{dsf}) becomes
\begin{eqnarray*} 
  - {\partial S_{pf} \over \partial {\mathcal A_{\mu}}(x) }
 \end{eqnarray*}
\begin{eqnarray*} 
  =(\ref{DAE})+ \lbrack(\ref{DAE})+(\ref{DAO})
  \rbrack(X^{A} \to X, Y^{A} \to Y) +
   (\ref{DB}), ~ ~ if ~ x=even,
 \end{eqnarray*}
\begin{eqnarray} 
  = (\ref{DAO})  + \lbrack(\ref{DAE})+(\ref{DAO})
  \rbrack(X^{A} \to X, Y^{A} \to Y) +
(\ref{DB}), ~ ~ if ~ x=odd.
\label{sum}
\end{eqnarray}
Note that the difference in the form of
the molecular dynamics equation on
even and odd sites is
only in the first term.

  \section{Discussions}
In this paper,
I have derived  
 (\ref{dsg}) and
 (\ref{sum}), relevant for equation of motion (\ref{mol})
 in molecular dynamics (or HMC)
  simulations  with clover
  fermions.
  I have also  extended the even-odd precondition technique, previously
  introduced for the
  quark
  propagator measurements, to the case of simulations
  with dynamical clover
  fermions.
  With the preconditioning technique,
  the number of iterations required would be reduced by a factor of 
  $3$ according to experience, and
  the most expensive part of the
  fermionic inversion is performed only on
  half
  lattice size. Therefore, it is expected
  that the preconditioned equation
  of motion would lead to considerable
 improvement over the unpreconditioned one.
  This scheme is  vectorizable and has been parallelized.
  Currently, the simulations of QCD at finite temperature
  using the clover action  are being carrying out
  on the Quadrics-APE100
  parallel computers. The above work might lay a foundation of further
  computer simulations using dynamical clover fermions.
  
  Of course, to obtain physical results from the numerical simulations,
  there are still
  of lot of work to do. For instance, because the clover constant
  $C$ depends dynamically
  on the gauge configuration, then there would be delicate interplay
  between $C$ and
  $\beta$, $\kappa$, in particular when the system is at criticality.
 This is a new subject beyond the scope of the paper.

   It has been mentioned in \cite{UKQCD} that even for
  the quenched clover propagator calculations, each minimum residue
  iteration took $35 \%$ longer than for the Wilson action.
  One has to choose a more efficient algorithm for fermionic inversion
  because a good algorithm is critical
  for full simulations.
  It is known that the
  stabilized biconjugate gradient is more efficient
  than the minimum residue or conjugate gradient,
  reducing the \rm{CPU} time
  by at least a factor of 1.5.
  
  Even if with the 
  clover action there is an improvement of the 
  finite cut-off error, the simulations with this action 
  require larger statistical samples, more arithmetic operations
  and much memory. 
  Concerning the feasibility of a full 
  QCD clover simulation on supercomputers,
  it is not easy to quantify, because it is machine and code dependent.
  
  I hope to discuss these problems and
  report the physical results in the near
  future.

\ack
  
I am grateful to R. Horsley for valuable discussions
on TAO (the programming
language of Quadrics-APE100), P. Lepage
on his improved perturbation theory,
H. Shanahan for reference \cite{Stanford},
and D. Richards and some members of
UKQCD collaboration for useful conversations about  $A=LDL^{\dagger}$
decomposition at Lattice 95.  
This work is sponsored by DESY.

\end{document}